\documentclass{iau}
\usepackage{graphicx}
\usepackage{natbib}
\usepackage{color, colortbl}
\definecolor{Gray}{gray}{0.9}

\def\msun{{\rm M}_\odot}

\title[Formation of millisecond pulsars via accretion with magnetic field] %% give here short title %%
{Formation of millisecond pulsars - NS initial mass and EOS constraints}

\author[Bejger et al.]   %% give here short author list %%
{Micha{\l}  Bejger$^1$, Morgane Fortin$^{1,2}$, Pawe{\l} Haensel$^1$ \\ \and J. Leszek Zdunik$^1$}

\affiliation{$^1$N. Copernicus Astronomical Center PAS, Bartycka 18, PL-00-716 Warsaw, Poland\\
email: {\tt bejger, haensel, jlz@camk.edu.pl } \\[\affilskip]
$^2$LUTH, UMR 8102 du CNRS, Observatoire de Paris, F-92195 Meudon Cedex, France\\
email: {\tt morgane.fortin@obspm.fr}}

\pubyear{2012}
\volume{290}  %% insert here IAU Symposium No.
\jname{Feeding compact objects: Accretion on all scales}
\editors{C.M. Zhang, T. Belloni, M. M\'endez \& S.N. Zhang, eds.}

\begin{document}
\maketitle
\begin{abstract}
Recent measurement of a high millisecond pulsar mass (PSR J1614-2230, $1.97\pm
0.04\ M_\odot$) compared with the low mass of PSR J0751+1807 ($1.26\pm 0.14\
M_\odot$) indicates a large span of masses of recycled pulsars and suggests a broad
range of neutron stars masses at birth. We aim at reconstructing the pre-accretion
masses for these pulsars while taking into account interaction of the magnetic
field with a thin accretion disk, magnetic field decay and relativistic 2D
solutions for stellar configurations for a set of equations of state. We briefly
discuss the evolutionary scenarios leading to the formation of these neutron stars
and study the influence of the equation of state.

\keywords{stars: neutron, pulsars: individual, accretion disks, equation of state,
magnetic fields}
%% add here a maximum of 10 keywords, to be taken form the file <Keywords.txt>
\end{abstract}
%----------------------------------------------------------------------
\firstsection
\firstsection
\section{Introduction}
%----------------------------------------------------------------------
The discovery of a $1.97\pm0.04\ M_\odot$ rapidly-rotating neutron star (NS)
\citep{Demorest2010} motivates the theorists to understand its origin and
composition. Population of millisecond pulsars contains also a low-mass 
PSR J0751+1807 ($1.26\pm 0.14\ M_\odot$, \citealt{NiceSK2008}); 
a statistical analysis of the measured and estimated pulsar
masses, in particular of the millisecond ones, was presented in \citep{Zhang2011}.
Despite a striking difference in their masses, their spin periods and inferred
magnetic fields are similar ($B\propto \sqrt{P\dot{P}}$, see Table~\ref{tab1}).
Both of them were {\it spun-up} by accretion from a binary companion. The evolution
of such systems is typically composed of several stages: starting from the Zero Age
Main Sequence (ZAMS), more massive primary (NS progenitor) enters the Red Giant
(RG) phase and engulfs the secondary for a brief Common Envelope (CE) phase.
Subsequently the primary, now as a helium star of a substantially lower mass,
explodes as a core-collapse supernova (SN), forming the NS. Secondary star evolves
from a ZAMS star into a RG and spins-up the NS via the Roche lobe overflow disk
accretion. Depending on stellar masses, the resulting composition of the secondary,
binary orbital periods and semi-major axes may be very different, which in turn may
result in additional stages of evolution: Intermediate Mass X-ray Binary (IMXB)
stage, connected with the Terminal Age Main Sequence instabilities in the case of
PSR J1914-2230, or secondary evaporation in a Low Mass X-ray Binary (LMXB) -
so-called Black Widow scenario - in the case of PSR J0751+1807 (see
Table~\ref{tab2} for quantitative figures and \citealt{DeLooreD1992},
\citealt{TaurisH2006} and \citealt{TaurisLK2011} for more details).
%----------------------------------------------------------------------
\begin{table}[t]
\begin{center}
\caption{Selected parameters of millisecond pulsars PSR J1614-2230 and
PSR J0751+1807 and their binary systems (orbital period $P_{\rm b}$, eccentricity
$e$ and the companion mass $M_{\rm WD}$; mass measurement errors reflect 1-$\sigma$ 
confidence levels).}
\begin{tabular}[t]{ccccccc}
\hline
PSR &$M\ [M_\odot]$&$P\ [{\rm ms}]$&$\dot{P}\ [10^{-21}\ s]$&
$P_{\rm b}\ [{\rm d}]$&$e$ &$M_{\rm WD}\ [M_\odot]$\\
 \hline
J1614-2230 & $1.97\pm 0.04$ & 3.15 & 9.6  &  8.7 d
     &$1.3\times 10^{-6}$ & $0.500\pm 0.006$ (CO WD) \\
J0751+1807 & $1.26\pm 0.14$ & 3.48 & 7.2  &  6 h
     &$5.6\times 10^{-5}$ & $0.12\pm 0.02$ (He WD)\\
 \hline
\label{tab1}
\end{tabular}
\end{center}
\end{table}
%----------------------------------------------------------------------
%----------------------------------------------------------------------
\newcolumntype{g}{>{\columncolor{Gray}}c}
\begin{table}[b]
\begin{center}
\caption{Evolutionary stages for PSR J1614-2230 and PSR J0751+1807 systems.}
\begin{tabular}[t]{ccc|g|cc|g|cc}
\hline
  &  & ZAMS & 1st RG & CE &SNII&2nd RG&now \\
 \hline
J1614-2230 & primary  & $25\;\msun$  &  $5\times 10^6\;$y & $7\;\msun$
& $1.9\;\msun$ &   &  $1.97\;\msun$ \\
  &secondary  &  $4.5\;\msun$ &   &  $4.5\;\msun$   & $4.5\;\msun$
  &  $5\times 10^7\;$y & $0.50\;\msun$ \\
  & $P_{\rm orb}$ & $1\;$y  &    &  $4\;$d  &  $2\;$d &    &  $8.7\;$d  \\
\hline
J0751+1807 & primary  & $15\;\msun$  &  $10^7\;$y & $5\;\msun$
& $1.1\;\msun$ &   &  $1.26\;\msun$ \\
  &secondary  &  $1.6\;\msun$ &   &  $1.6\;\msun$   & $1.6\;\msun$
  &  $10^9\;$y & $0.12\;\msun$ \\
  & $P_{\rm orb}$ & $1\;$y  &    &  $1\;$d  &  $1.5\;$d &    &  $6\;$h\\
 \hline
\label{tab2}
\end{tabular}
\end{center}
\end{table}
%----------------------------------------------------------------------
%----------------------------------------------------------------------
\firstsection
\section{Methods}
%----------------------------------------------------------------------
Hereunder we focus on the recycling/spin-up stage of the evolution leading to the
formation of a millisecond pulsar. It is modeled by a sequence of stationary
rotating configurations of increasing mass, calculated for a given equation of
state (EOS). Rigidly rotating axisymmetric stellar configuration were obtained by
means of the library for relativistic computations {\tt LORENE}. 
Increase of the total stellar angular momentum $J$ is calculated by accounting for the transfer of
the angular momentum from the thin accretion disk interacting with the NS
magnetosphere:
\begin{equation}
\frac{{\rm d}J}{{\rm d}M_{\rm b}}=l_0 - \frac{K_B}{\dot M},
\label{eq:evol}
\end{equation}
where an accreted (baryon) mass ${\rm d}M_b$ carries a specific angular momentum
$l_0$ from the inner boundary of a disk; $\dot M$ is the accretion rate and $K_B$
denotes a magnetic (braking) torque. We employ the prescription of
\cite{KluzniakR2007}, improved by considering the relativistic effects (the
innermost stable circular orbit, \citealt{BejgerFHZ2011}) and using approximate,
but accurate representation for orbital parameters of particles in the disk
\citep{BejgerZH2010}. Gradual magnetic field decay, which is caused by the
infalling accreted matter, is taken into account as well; we adopt the relation of
\citet{Shibazaki1989} and \citet{TaamH1986} that links the decrease of $\vec{B}$
with the amount of accreted matter only (actual evolution of the polar magnetic
field of an accreting NS can be more complicated, see \citealt{ZhangKojima2006} and
references therein).

Following EOSs were used: APR by \citet{AkmalPR1998}, DH by
\citet{DouchinH2001} and BM by \citet{BednarekHZBM2011}. APR EOS (${\rm
A}18+\delta v+{\rm UIX}^\ast$ model) is a variational, non-relativistic many-body
solution with relativistic corrections (maximal mass of a non-rotating star:
$M^{\rm stat}_{\rm max}=2.19\ M_\odot$, $R(M^{\rm stat}_{\rm max})=9.93$ km).  DH
EOS is constructed using a non-relativistic energy density functional based on the
SLy4 effective nuclear interaction, designed to describe both crust and core in an
unified way ($M^{\rm stat}_{\rm max}=2.05\ M_\odot$, $R(M^{\rm stat}_{\rm
max})=10.0$ km).  BM EOS is based on relativistic mean field model, with a
non-linear Lagrangian that includes quartic terms in meson fields, and two
additional hidden-strangeness mesons $\sigma^*$ and $\phi$ that couple to hyperons
only ($M^{\rm stat}_{\rm max}=2.03\ M_\odot$, $R(M^{\rm stat}_{\rm max})=10.7$ km).

Detailed description of methods and tests can be found in \citet{BejgerFHZ2011}.
%----------------------------------------------------------------------
\firstsection
\section{Results}
\begin{figure}[h]
\begin{center}
\includegraphics[width=2.625in]{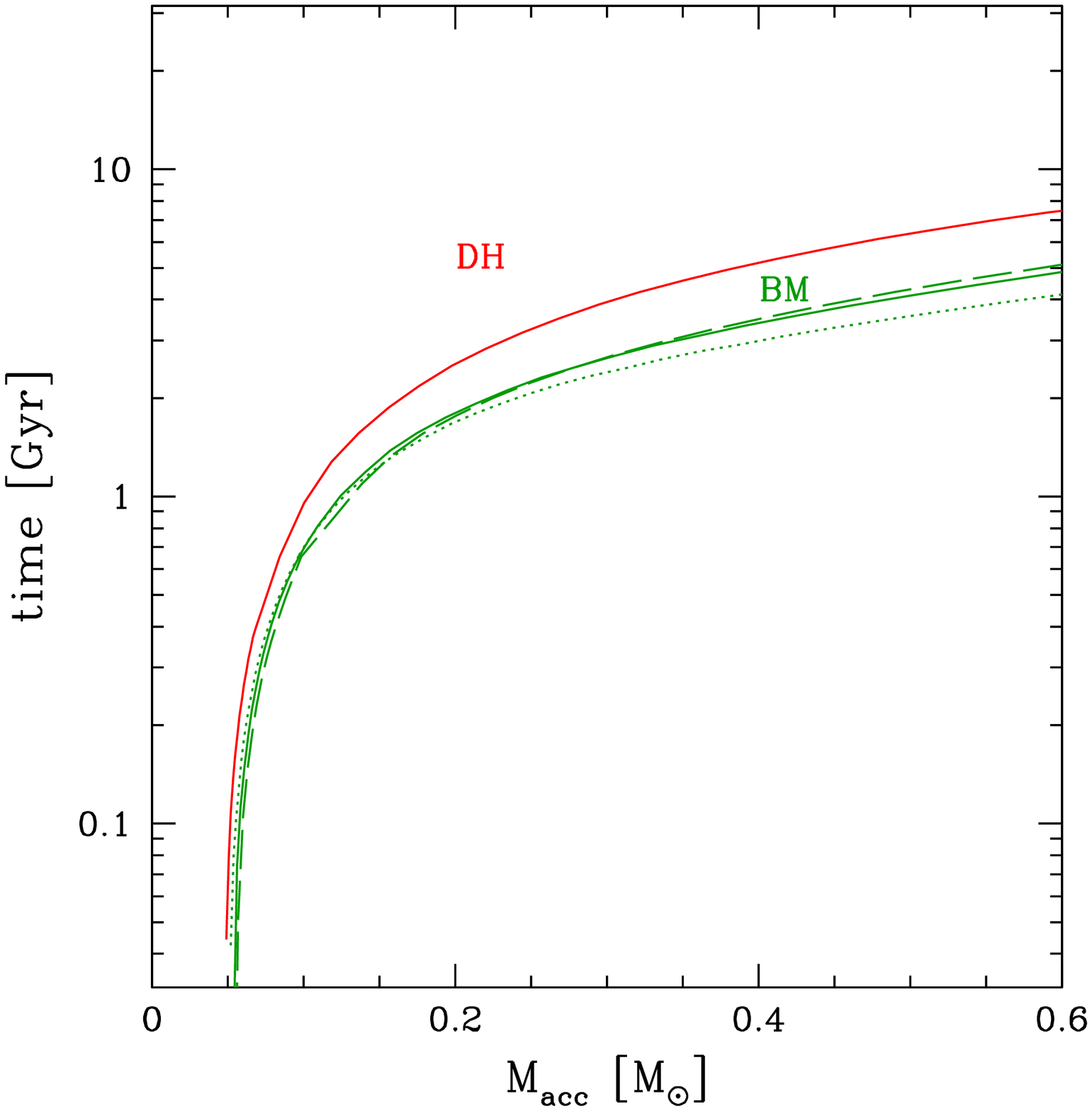}
\includegraphics[width=2.625in]{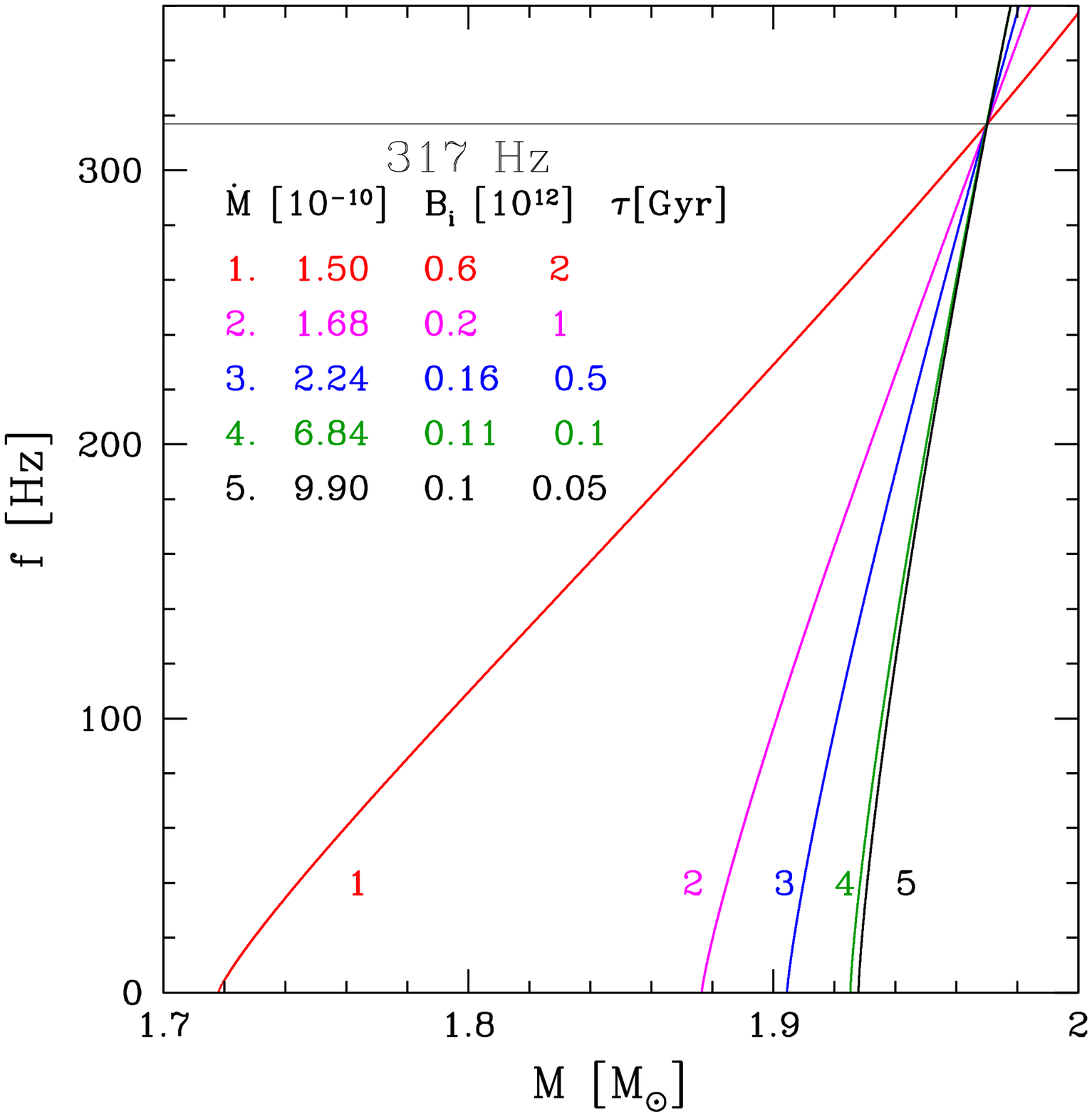}
\caption{Currently most massive millisecond pulsar PSR J1614-2230. Left panel:
Accreting time needed to spin-up the star to presently observed values as
a function of accreted mass for DH and BM EOSs ($B_i=10^{12}$ G, dotted and dashed
lines correspond to lower and upper mass limit, $1.93$ and $2.01\ M_\odot$).
Right panel: selected spin-up tracks parametrized by different initial masses,
magnetic fields, accretion rates and times. Lines cross at the current
values of the pulsar parameters (BM EOS).}
   \label{fig1}
\end{center}
\end{figure}

\begin{figure}[h]
\begin{center}
\includegraphics[width=2.625in]{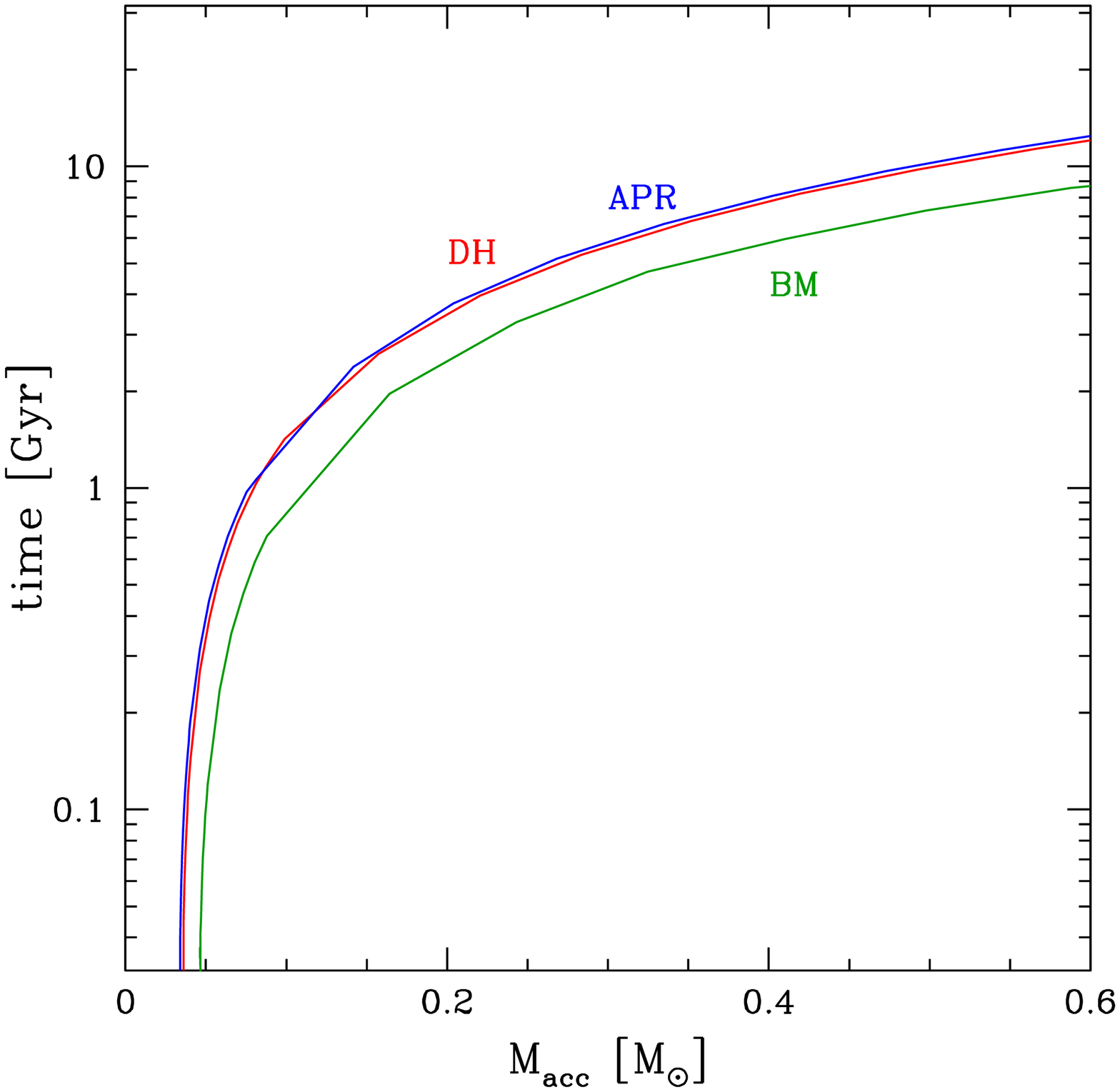}
\includegraphics[width=2.625in]{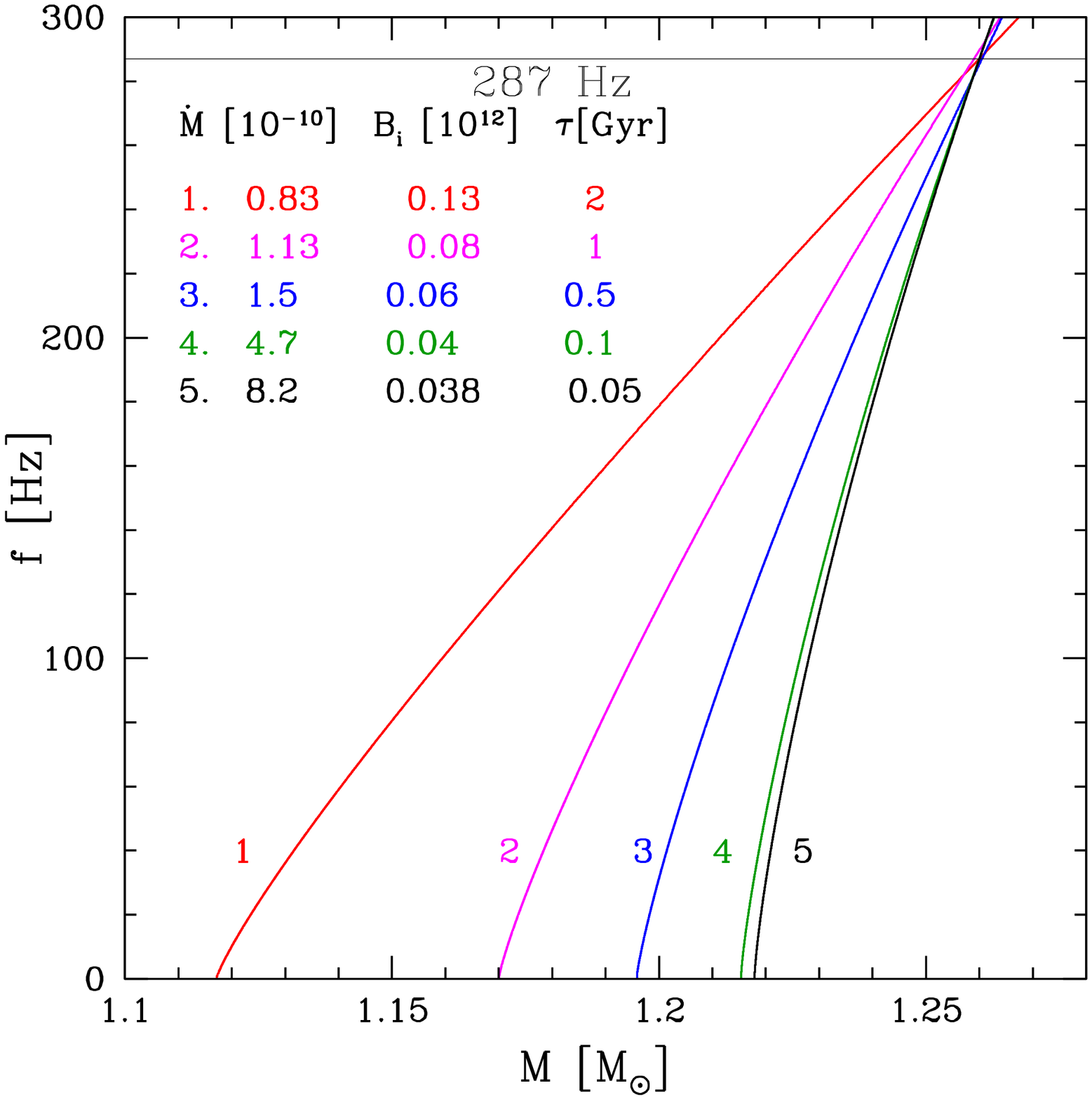}
\caption{As in Figure 1, but for a low-mass millisecond pulsar PSR J0751+1807.}
   \label{fig2}
\end{center}
\end{figure}

%----------------------------------------------------------------------
The recycling process for PSR J1614-2230 and J0751+1807 is presented in
Fig.~\ref{fig1} and \ref{fig2}, respectively, and visualized in terms of accretion
time $\tau$, the amount of accreted mass $M_{\rm acc}$,  and stellar spin frequency
$f$. We assume constant accretion rate $\dot{M}$ during the recycling. The lifetime
of an LMXB or IMXB system puts strong limits on the amount of accreted mass and
$\dot M$. In the case of J1614-2230, such an evolutionary argument of
\citet{TaurisLK2011} places a constraint on the accretion time $\tau$: $\tau \simeq
100\ {\rm Myr}$, which results in a pre-accretion mass of this NS very close to the
observed mass: $> 1.9\ M_\odot$, and a way to estimate $\dot M$ needed for the
spin-up (see second term in Eq.~\ref{eq:evol}; small $\tau$ requires large $\dot M$
close to the Eddington limit).

From Fig.~\ref{fig1} and \ref{fig2} one can deduce the minimal amount of
accreted mass needed to reach the observed $\simeq 300$ Hz spin frequency:
$\Delta M_{\rm min} \simeq 0.04 \div 0.05\ M_\odot$ for $\tau < 1$ Gyr. Hence, 
$\dot M$ in such a case is 
\begin{equation}
 \dot M=\frac{\Delta M_{\rm min}}{\tau}\simeq 10^{-9}\ M_\odot/{\rm yr}\ {\tau_{50}}^{-1},
\end{equation}
where ${\tau_{50}}=\tau/50$ Myr. For PSR J0751+1807, our simulations indicate an
initial mass of $\simeq 1.2 M_\odot$; its value is a function the accretion time $\tau$,
$\dot M$ and initial magnetic field (see right panel in Fig.~\ref{fig2}).

The influence of the dense-matter EOS can be summarized in terms of its stiffness:
the stiffer the EOS, the larger the  moment of inertia $I$ for a given mass ($I\sim
MR^2$); it requires therefore more accretion of angular momentum to spin-up a star
to a given frequency in a given amount of time (see Eq.~\ref{eq:evol}). While the
DH and APR EOSs give very similar results for observed frequencies, the BM EOS
differs from them since it must be much stiffer ($I$ is $20\%$ higher than for DH
EOS) to sustain the softening by  hyperons (see e.g., left panel on
Fig.~\ref{fig2}).

%----------------------------------------------------------------------
%----------------------------------------------------------------------
%----------------------------------------------------------------------
%----------------------------------------------------------------------
%----------------------------------------------------------------------
\firstsection
\section{Summary}
%----------------------------------------------------------------------
We present a versatile method of studying the formation and evolution of
millisecond pulsars in the recycling phase. This prescription allows for testing
the magnetic field decay and magnetic torque models,   in the framework of a
relativistic thin accretion disk spin-up scenario. Coupled with evolutionary data,
the information relevant to the SN collapse simulations, spin and initial mass distribution of
pulsar populations can be drawn from the timing data and mass measurements. We
point out that from the dense-matter EOS point of view, both the low and the high
end of the initial mass spectrum is very important, as both the minimal and the
maximal initial NS mass contains the information about the composition of the
matter and processes occurring at its birth. Moreover, the disk--magnetic field
interaction in our model changes dramatically the character of the evolution of the
interior and the spin of NS, as compared to the results where $\vec{B}$ is
neglected \citep{BejgerHZF2011}. 
\vskip 0.1cm 
{\it Acknowledgements.} This work was partially supported by 
the Polish MNiSW research grant no. 2011/01/B/ST9/04838.
\firstsection

\end{document}